\documentclass[aps,prl,showpacs,twocolumn]{revtex4}
\usepackage[utf8]{inputenc}
\usepackage{amsmath}
\usepackage{amssymb}
\usepackage{amsthm}
\usepackage{graphicx}
\usepackage{graphics}
\begin{document}

\title{Bulk Antisymmetric tensor fields coupled to a dilaton in a
Randall-Sundrum model}

\author{G. Alencar$^1$, M. O. Tahim$^1$, R. R. Landim$^2$, C. R. Muniz$^{3}$ and R. N. Costa Filho$^{2,4}$}

\affiliation{$^1$ Universidade
Estadual do Cear\'a, Faculdade de Educa\c c\~ao, Ci\^encias e Letras do Sert\~ao Central, Quixad\'a,Cear\'a, Brazil.\\$^2$Departamento de F\'{\i}sica, Universidade Federal do Cear\'{a}, Caixa
Postal 6030, Campus do Pici, 60455-760, Fortaleza, Cear\'{a}, Brazil.\\$^3$Universidade
Estadual do Cear\'a, Faculdade de Educa\c c\~ao, Ci\^encias e Letras de Iguatu, Rua
Deocleciano Lima Verde, s/n Iguatu,Cear\'a, Brazil.\\$^4$Department of Physics and
Astronomy,University of Western Ontario London, Ontario, Canada N6A 3K7.}

\date{\today}

\begin{abstract}

A string-inspired 3-form-dilaton-gravity model is studied in a Randall-Sundrum brane world scenario.
As expected, the rank-3 antisymmetric field is exponentially suppressed. For each mass level, the mass
spectrum is bigger than the one for the Kalb-Ramond field. The coupling between the dilaton and the massless
Kaluza-Klein mode of the 3-form is calculated and the coupling constant of the cubic interactions obtained
 numerically. This coupling are of the order of Tev$^{-1}$, therefore there exist a possibility to find some
signal of it at Tev scale.

\vspace{0.75cm}

\end{abstract}


\maketitle

\section{Introduction}

The core idea of extra dimensions models is to
consider the four-dimensional universe as a hyper-surface embedded in
multidimensional manifold. The appeal of such models is the determination of
scenarios where membranes have the best chances to mimic the standard model's
characteristics. In particular, the standard model presents interesting topics to
study such as the hierarchy problem, and the cosmological constant problem that can be
treated by the above-mentioned models. For example, the Randall-Sundrum model \cite{Randall:1999vf,Randall:1999ee}
provides a possible solution to the hierarchy problem and show how gravity is trapped to a membrane.

There are studies which concern to general properties of membranes, for example,
one that makes an analysis of the space-time singularities that arise in braneworld models
with 4D Poincare invariance \cite{Herrera-Aguilar}.
Others with cosmological implications, that
consider our universe initially as an empty p-brane embedded in a manifold, whose bulk is populated
only with a dilaton field that couples itself to the brane,
producing a rich and interesting evolutionary cosmology \cite{DeRisi:2006pz}.
The presence of other fields in the bulk rises the problem
of field localization in the brane,
that is an important tool in order to build up the standard model in
the membrane setup. For such, several ingredients had to be added:
the gravitational field, spinor fields, scalar fields and gauge
fields. A lot of aspects have been studied related to these topics.
These investigations include both the smooth models described by
soliton-like membranes and the pure Randall-Sundrum models
containing space-time singularities. For these there is an extensive
list of references in literature
\cite{Liang:2009zz,Liang:2009zze,Liu:2009uca,Flachi:2009uq,Koley:2008tn,Bazeia:2007nd,Liu:2007ku,BarbosaCendejas:2006hj,
Midodashvili:2003ib,Kehagias:2000au,Liu:2010pj,Davies:2007tq}.

Despite its achievements, that model does not favor the localization of zero modes of gauge vector fields.
Such localization is important to construct the standard model in a membrane \cite{Davoudiasl:1999tf}.
That happens because the gauge vector field theory is conformal in $D=4$: this makes the warp factor in the
pure Randall-Sundrum model disappear from the effective action.
The issue of localization of standard model fields
on the three-brane has been widely studied \cite{Rubakov:1983bb,Dvali:1996xe,Pomarol:1999ad,Grossman:1999ra,ArkaniHamed:1999dz}. Despite of
this, we must consider a scenario in witch it is assumed \textit{a priori} that the matter fields are constrained to live on the three-brane.

Some studies about higher rank tensor fields have been made showing its
relation with the AdS/CFT conjecture \cite{Germani:2004jf}. Besides this,
String Theory shows the naturalness of higher rank tensor fields in its
spectrum\cite{Polchinski:1998rq,Polchinski:1998rr}.

In mathematical terms, the presence of one more extra dimension ($D=5$) provides the
existence of many antisymmetric fields, namely the two, three,
four and five forms. However, the only relevant ones for the
visible brane are the two and the three form. This is due to the
fact that when the number of dimensions increase, also increases the
number of gauge freedom. This can be used to cancel the dynamics of
the field in the visible brane. The mass spectrum of the two and
three form have been studied, for example, in Refs.
\cite{Mukhopadhyaya:2004cc} and \cite{Mukhopadhyaya:2007jn}.

The coupling between the two form and the dilaton, with cosmological
consequences, have been studied in \cite{DeRisi:2007dn}. There are also
authors that investigate that coupling in domain of standard model
physics \cite{Mukhopadhyaya:2009gp}.
Such coupling, inspired in string theory, can provide us with a
process that, in principle, could be observed in the LHC. That may
happen through a Drell-Yang process, in which a pair of
quark-antiquark can give rise to a three form field, mediated by a
dilaton. This kind of process can appear in a scenario where
the dilaton is considered as the Higgs field \cite{Ryskin:2009kw,Foot:2008tz,Goldberger:2007zk}.
This raises the question if the coupling between the three
form and the dilaton give a similar process, and this is the goal of
this article.

The question to be addressed here in this piece of work is related to other kind of gauge
fields.  We present a full action, including
the gravity, the dilaton and the three form field action. The
dimensional reduction is done and the effective action at the
visible brane is obtained. Then, we analyze
carefully the dilaton sector, find the mass spectrum and find the
complete solution for dilaton coefficients. This result will be
needed when the coupling with the three form is analyzed. After, we analyze the three form sector and, like in the
dilaton case, we find the mass spectrum and solve the equation to
find the coefficients of the expansion. The mass spectrum of the
three form is compared with that of the two form, the dilaton and
the gravity field. This gives a hint that the three form
must not be seen at LHC, because of the mass of the massive Kaluza-Klein modes.
Finally we study the coupling with the dilaton and find how we can conclude that a signal of the massless three form can be found at LHC.

\section{Dilaton and Antisymmetric Tensor Fields in Randall Sundrum Framework}

As the number of antisymmetric tensor fields increase with
dimensions, these fields has to be considered. In fact these fields
have been taken into account in the literature. The fact is that,
when the number of dimensions increases, the number of gauge freedom
also increases, and this can be used to cancel the degrees of
freedom in the visible brane. Therefore the antisymmetric tensors
relevant to the visible brane are that of rank two and three
\cite{Mukhopadhyaya:2007jn}. We focus our attention here in the
second tensor field cited. As usual, capital Latin index represent the coordinates
in the bulk and Greek index, those on the brane. The metric is given by
\begin{equation}
ds^{2}=e^{-2\sigma }\eta _{\mu \nu }dx^{\mu }dx^{\nu }+r_{c}^{2}dy^{2}
\end{equation}
where $\sigma =kr_{c}y$, $y$ is the coordinate of fifth dimension, $r_c$ is the compactification
radius for that  dimension, $k$ is a constant of the order of the higher dimensional Planck mass $M$
and $\eta _{\mu \nu }$ is the $4D$ Minkowski metric.

As said in the introduction, this is a string inspired model. The type IIA superstring posses the two and
the three form in the spectrum, the first coming from the NS-NS sector and the other form the RR \cite{Polchinski:1998rr}.
The dilaton determines the coupling constant of string theory, therefore the tree level space-time action has an overall factor of it.
This means that the coupling of the two form is the same as that of the three form in the effective action.
With these arguments at hand, we consider the same coupling as that used recently by Mukhopadhyaya,
where a two form is coupled to the dilaton \cite{Mukhopadhyaya:2009gp}. The action to be considered is therefore given by
\begin{align}
S=S_{grav}+S_{X}+S_{dil}
\end{align}
where
\[
S_{grav}=\int d^{5}x\sqrt{-G}2M^{3}R,
\]
\[
S_{X}=\int d^{5}x\sqrt{-G}\left[-e^{\phi /M^{3/2}}2Y_{MNLP}Y^{MNLP}\right]
\]
and
\begin{equation}
S_{dil}=\int d^{5}x\sqrt{-G}\left[\frac{1}{2}\partial _{M}\phi \partial ^{M}\phi
-m^{2}\phi ^{2}\right].
\end{equation}

Defining the $3-$form field $X_{MNO}$ and fixing the gauge $X_{\mu \nu y}=0$, we have two possibilities for the strength tensor
\[
Y_{\mu \nu \alpha \beta}=\partial _{\lbrack \mu }X_{\nu \alpha \beta ]}
\]
or
\[
Y_{y \nu \alpha \beta}=\partial _{\lbrack y }X_{\nu \alpha \beta]}
.\]

We must here focus only in the dilaton and the antisymmetric field
$X_{MNO}$. Thus, for the metric in $(1)$, we have
\begin{eqnarray*}
S_{X} &=&\int d^{4}x\int dyr_{c}e^{-4\sigma }\lbrace{A+B+C\rbrace},
\end{eqnarray*}
where
\begin{equation}
A=-2e^{\phi
/M^{3/2}}e^{8\sigma }\eta ^{\alpha \beta }\eta ^{\mu \nu }\eta ^{\lambda
\delta }\eta ^{\gamma \tau }Y_{\alpha \mu \lambda \gamma }Y_{\beta \nu
\delta \tau },
\end{equation}
\begin{equation}
B=-2e^{\phi
/M^{3/2}} \frac{4}{r_{c}^{2}}e^{6\phi }\eta ^{\alpha \beta }\eta ^{\mu
\nu }\eta ^{\lambda \gamma }\partial _{y}X_{\alpha \mu \lambda }\partial
_{y}X_{\beta \nu \gamma }
\end{equation}
and
\begin{equation}
C= r_{c}e^{-4\sigma }\left[\frac{e^{2\sigma }}{2}\eta ^{\mu \nu }\partial _{\mu
}\phi \partial _{\nu }\phi +\frac{1}{2r_{c}^{2}}\partial _{y}\phi \partial
_{y}\phi -m^{2}\phi ^{2}\right].
\end{equation}

As the mass term of the dilaton decouples the visible brane we must
consider the kinetic term only. Performing an integration by parts in the
second terms of the dilaton and $X$ fields respectively we get
\begin{align*}
S &=\int d^{4}x\int dy\left[r_{c}e^{-4\sigma }A\right.\\
&+e^{\phi /M^{3/2}}\frac{8}{r_{c}}\eta ^{\alpha
\beta }\eta ^{\mu \nu }\eta ^{\lambda \gamma }X_{\alpha \mu \lambda
}\partial _{y}\left( e^{2\phi }\partial _{y}X_{\beta \nu \gamma
}\right)  \\
&+\left.\frac{r_{c}e^{-2\sigma }}{2}\eta ^{\mu \nu }\partial _{\mu }\phi
\partial _{\nu }\phi -\frac{1}{2r_{c}}\phi \partial _{y}\left( e^{-4\sigma
}\partial _{y}\phi \right) \right].
\end{align*}

Note that the derivative in the exponential of the dilaton would
give a term suppressed by the Planck mass and therefore is
irrelevant for the effective action. At this point we must consider
the Kaluza-Klein decomposition of the fields
\begin{eqnarray*}
X_{\mu \nu \alpha } &=&\sum_{n=0}^{\infty }X_{\mu \nu \alpha }^{n}(x)\frac{%
\chi ^{n}(y)}{\sqrt{r_{c}}} \\
\phi &=&\sum_{n=0}^{\infty }\phi ^{n}\left( x\right) \frac{\psi ^{n}\left(
y\right) }{\sqrt{r_{c}}}
\end{eqnarray*}%
where
\begin{eqnarray}
\int e^{4\sigma }\chi ^{m}\left( y\right) \chi ^{n}\left( y\right) dy
&=&\delta ^{mn},  \nonumber \\
\int e^{-2\sigma }\psi ^{m}\left( y\right) \psi ^{n}\left( y\right) dy
&=&\delta ^{mn}.  \label{Orthonormality}
\end{eqnarray}

In terms of the above projections, the effective action on the
visible brane is given by
\begin{eqnarray*}
S &=&\int d^{4}x\sum \limits_{n=0}^{\infty
}\sum\limits_{m=0}^{\infty }\int dy\{e^{\phi /M^{3/2}}\\
&&\times[-2e^{4\sigma
}\chi ^{m}\chi ^{n}\eta ^{\alpha \beta }\eta ^{\mu \nu }\eta
^{\lambda \delta }\eta ^{\gamma \tau }Y_{\alpha \mu \lambda \gamma
}^{n}Y_{\beta \nu \delta \tau }^{m}\\
&&+\frac{8}{r_{c}^{2}}\eta ^{\alpha
\beta }\eta ^{\mu \nu }\eta ^{\lambda \gamma }X_{\alpha \mu \lambda
}^{n}X_{\beta \nu \gamma }^{m}\chi ^{m}\frac{d}{dy}\left( e^{2\phi }
\frac{d}{dy}\chi ^{n}\right) ] \\
&&+[\frac{e^{-2\sigma }}{2}\psi ^{m}\psi ^{n}\eta ^{\mu \nu }\partial _{\mu
}\phi ^{m}\partial _{\nu }\phi ^{n} \\
&&-\frac{1}{2r_{c}^{2}}\phi ^{m}\phi^{n}\psi ^{m}\partial _{y}\left( e^{-4\sigma }\partial _{y}\psi ^{n}\right)
]\}.
\end{eqnarray*}

The action above is the action of the rank three antisymmetric
tensor field, or three form, coupled to a dilaton. The first bracket
represents the three form coupled to a dilaton and the second the
dilaton itself. As there is no coupling of the terms in the second
brackets and the three form fields, it is a free dilaton action. So
we must concentrate in the free dilaton action and later go on to
the main objective, which is the antisymmetric field.

\section{Bulk Dilaton Field}

At this section we must consider the dilaton sector. Here we must
expand the dilaton exponential
\[
e^{\phi /M^{3/2}}=1+\left( \phi /M^{3/2}\right) +\left( \phi /M^{3/2}\right)
^{2}/2!+...
\]
and concentrate only in the first term which will give the usual
kinetic and mass term for the dilaton field provided we have
\begin{equation}
-\frac{1}{r_{c}^{2}}\partial _{y}\left( e^{-4\sigma }\partial _{y}\psi
^{n}\right) =\left( m_{n}^{d}\right) ^{2}e^{-2\sigma }\psi ^{n}.
\label{eq massa dil}
\end{equation}

The other terms of the expansion will be considered later as
interaction terms in the effective action. Using the above orthonormality conditions, we
finally get
\[
S_{dil}=\int d^{4}x\sum\limits_{n=0}^{\infty }\frac{1}{2}[\eta ^{\mu
\nu }\partial _{\mu }\phi ^{n}\partial _{\nu }\phi ^{n}+\left(
m_{n}^{d}\right) ^{2}\phi ^{n}\phi ^{n}].
\]

Therefore we obtain a standard dilaton action with masses given by
the solutions of equation (\ref{eq massa dil}). The solution is
given in \cite{Mukhopadhyaya:2009gp} and we will basically repeat it
here. The easiest solution is for the massless dilaton, where we
have to solve the equation
\[
-\frac{1}{r_{c}^{2}}\partial _{y}\left( e^{-4\sigma }\partial _{y}\psi
^{n}\right) =0.
\]
with obvious solution
\[
\psi ^{0}=\frac{C_{1}}{4kr_{c}}e^{4\sigma }+C_{2}.
\]

The constants can be obtained from orbifold condition $\psi^{0}(-\pi)=\psi^{0}(\pi)$ and from
orthonormality conditions, again. We find $C_{1}=0$ and, considering also that $\exp{(-2kr_{c}\pi)}<<1$, $C_{2}=\sqrt{kr_{c}}$.
The final solution is
\[
\psi ^{0}=\sqrt{kr_{c}}.
\]

In order to solve the equations for the massive modes first we must make the change of variables $%
z_{n}=\frac{m_{n}^{d}}{k}e^{\sigma }$ and $f^{n}=\psi ^{n}e^{-2\sigma }$ to
obtain the second order equation
\[
\lbrack z_{n}^{2}\frac{d^{2}}{dz_{n}^{2}}+z_{n}\frac{d}{dz_{n}}%
+(z_{n}^{2}-4)]f_{n}=0,
\]%
which admits a Bessel function of order 2 as a solution. \ The
general solution is
\[
\psi ^{n}=\frac{e^{2\sigma }}{N_{n}}[J_{2}\left( z_{n}\right) +\alpha
_{n}Y_{2}\left( z_{n}\right) ]
\]
where $N_{n}$ and $\alpha _{n}$ are constants to be determined. First of all
we must to use the continuity condition for the derivative of $\psi ^{n}$ at
$y=0$. Remembering that $J_{2}$ and $Y_{2}$ are Bessel and Neumann functions
of order 2 we obtain
\[
\alpha _{n}=-\frac{J_{1}\left( \frac{m_{n}^{d}}{k}\right) }{Y_{1}\left(
\frac{m_{n}^{d}}{k}\right) }
\]

As the masses $m_{n}^{d}$ are expected to be of order of Tev on the brane we
have $m_{n}^{d}<<k$ and expanding the rhs of the above equation we obtain
\[
\alpha _{n}=-\frac{\pi }{4}\left( \frac{m_{n}^{d}}{k}\right)
^{2}<<1.
\]

Using the boundary condition at $y=\pi $ we obtain
\begin{equation}
J_{1}\left( x_{n}\right) =0  \label{dilaton mass}
\end{equation}%
where we have defined $x_{n}=z_{n}\left( \pi \right) =\frac{m_{n}^{d}}{k}%
e^{kr_{c}\pi }$.

If we find the solutions of the equation above we can find $x_{n}$
and therefore the mass spectrum. Now the orthogonality condition can
be used to determine $N_{n}$ and we obtain for the solution
\[
\psi ^{n}\left( z_{n}\right) =\sqrt{kr_{c}}\frac{e^{2\sigma }}{e^{kr_{c}\pi }%
}\frac{J_{2}\left( z_{n}\right) }{J_{2}\left( x_{n}\right) }.
\]

Now we have the complete solution for the dilaton given by
\[
\phi =\sqrt{kr_{c}}\phi ^{0}+\sum_{n=1}^{\infty }\sqrt{k}\frac{e^{2\sigma }}{%
e^{kr_{c}\pi }}\frac{J_{2}\left( z_{n}\right) }{J_{2}\left( x_{n}\right) }%
\phi ^{n}
\]
with masses given by equation (\ref{dilaton mass}). Posteriorly we will
compare the mass of the dilaton with the graviton one. The complete solution
for the dilaton will be necessary for studying its coupling to the
antisymmetric field.

\section{Bulk three form field}

Now we must consider the three form field. The action was constructed before
and we have
\begin{align}
S_{X}&=\int d^{4}x\sum\limits_{n=0}^{\infty
}\sum\limits_{m=0}^{\infty }\int dye^{\phi /M^{3/2}}\nonumber\\
&[-2e^{4\sigma
}\chi ^{m}\chi ^{n}\eta ^{\alpha \beta }\eta ^{\mu \nu }\eta
^{\lambda \delta }\eta ^{\gamma \tau }Y_{\alpha
\mu \lambda \gamma }^{n}Y_{\beta \nu \delta \tau }^{m}\nonumber\\
&+\frac{8}{r_{c}^{2}}\eta ^{\alpha \beta }\eta ^{\mu \nu }\eta ^{\lambda \gamma }X_{\alpha \mu
\lambda }^{n}X_{\beta \nu \gamma }^{m}\chi ^{m}\frac{d}{dy}\left(
e^{2\phi }\frac{d}{dy }\chi ^{n}\right) ].
\end{align}

Again we must expand of the dilaton exponential and consider only the first
term. The mass spectrum is obtained from
\[
-\frac{1}{r_{c}^{2}}\frac{d}{dy}\left( e^{2\phi }\frac{d}{dy}\chi
^{n}\right) =\left( m_{X}^{n}\right) ^{2}\chi ^{n}e^{4\sigma }
\]
and again the other terms of the expansion will be considered later as
interaction terms in the effective action. The simplest and most important
solution of the above equation is obtained for the massless case, where we
have the equation
\[
-\frac{1}{r_{c}^{2}}\frac{d}{dy}\left( e^{2\phi }\frac{d}{dy}\chi
^{n}\right) =0
\]%
with obvious solution
\[
\chi ^{0}=-\frac{C_{1}}{2kr_{c}}e^{-2\sigma }+C_{2}.
\]

Using continuity of the derivative at $y=\pi $ give us $C_{1}=0$ and
from orthogonality we finally get
\[
\chi ^{0}=\sqrt{2kr_{c}}e^{-2kr_{c}\pi }.
\]

For the massive modes, similarly to the dilaton case, we perform the
redefinitions $z_{n}^{\prime }=\frac{m_{n}}{k}e^{\sigma }$ and $%
f_{n}^{\prime }=e^{\sigma }\chi ^{n}$ to obtain the equation
\[
\lbrack z_{n}^{\prime 2}\frac{d^{2}}{dz_{n}^{\prime 2}}+z_{n}^{\prime }\frac{%
d}{dz_{n}^{\prime }}+\left( z_{n}^{\prime 2}-1\right) ]f_{n}^{\prime }=0
\]
with solution given by a Bessel function of order 1. Therefore
\[
\chi ^{n}=e^{-\sigma }f_{n}^{\prime }=\frac{e^{-\sigma }}{N_{n}^{\prime }}%
\left[ J_{1}\left( z_{n}^{\prime }\right) +\alpha _{n}^{\prime }Y_{1}\left(
z_{n}^{\prime }\right) \right] ,
\]%
and again we have to determine the constants $N_{n}^{\prime }$ and
$\alpha _{n}^{\prime }$ using contour conditions. First of all we
must use continuity conditions at $y=0$ to obtain
\[
\alpha _{n}^{\prime }=-\frac{J_{2}\left( \frac{m_{X}^{n}}{k}\right) }{%
Y_{2}\left( \frac{m_{X}^{n}}{k}\right) }.
\]

As the masses are in Tev scale we have $m_{X}^{n}<<k$ and expanding the
above expression we get
\[
\alpha _{n}^{\prime }\sim \frac{\pi }{2^{5}}\left( \frac{m_{X}^{n}}{k}%
\right) ^{4}<<1.
\]

Using the above result, the fact that $e^{kr_{c}\pi }>>1$ and the contour
condition at $y=\pi $ we obtain
\begin{equation}
J_{2}\left( x_{n}^{\prime }\right) =0  \label{mass eq X}
\end{equation}%
with the definition $x_{n}^{\prime }=z_{n}^{\prime }\left( \pi \right) =%
\frac{m_{X}^{n}}{k}e^{kr_{c}\pi }$. Therefore we can obtain the mass
spectrum and, in the visible brane, the masses are solutions of the equation
(\ref{mass eq X}). The effective action is given by
\begin{align}
S_{X}&=\int d^{4}x\sum\limits_{n=0}^{\infty }e^{\phi
/M^{3/2}}\{-2\eta ^{\alpha \beta }\eta ^{\mu \nu }\eta ^{\lambda
\delta }\eta ^{\gamma \tau }Y_{\alpha \mu \lambda \gamma
}^{n}Y_{\beta \nu \delta \tau }^{m}\nonumber\\
&-8\left( m_{X}^{n}\right)
^{2}\eta ^{\alpha \beta }\eta ^{\mu \nu }\eta ^{\lambda \gamma
}X_{\alpha \mu \lambda }^{n}X_{\beta \nu \gamma }^{m}\}.
\end{align}

We can obtain $N_{n}$ with the normalization condition to obtain
\[
N_{n}^{\prime }=\frac{e^{kr_{c}\pi }}{\sqrt{kr_{c}}}J_{1}\left(
x_{n}^{\prime }\right) ,
\]
and we arrive at the final solution for the massive modes given by
\[
\chi ^{n}\left( z_{n}\right) =\sqrt{kr_{c}}\frac{e^{\sigma }}{e^{kr_{c}\pi }}%
\frac{J_{1}\left( z_{n}^{\prime }\right) }{J_{1}\left( x_{n}^{\prime
}\right) }.
\]

It is interesting to note that, despite the way we have defined it,
the mass spectrum of the three form field is no altered by the
dilaton, and the only change will be in the interaction terms, that
will be analyzed posteriorly. We list in table \ref{masses} the value of the masses for
the graviton,the dilaton,the two and three form, as per \cite{Mukhopadhyaya:2007jn,Mukhopadhyaya:2009gp}, using the scale $kr_{c} = 12$ and $k = 10^{19}Gev$
\begin{table}[h]
\begin{center}
\begin{tabular}{|c|cccc|}
\hline
$n$ & 1 & 2 & 3 & 4 \\ \hline
$m_{grav}^{n}$(TeV) & 1.66 & 3.04 & 4.40 & 5.77 \\ \hline
$m_{dil}^{n}$(TeV) & 1.66 & 3.04 & 4.40 & 5.77 \\ \hline
$m_{KR}^{n}$(TeV) & 2.87 & 5.26 & 7.62 & 9.99 \\ \hline
$m_{X}^{n}$(TeV) & 4.44 & 7.28 & 10.05 & 12.79 \\ \hline
\end{tabular}
\end{center}
\caption{Masses of KK modes where $kr_{c} = 12$ and $k = 10^{19}Gev$}
\label{masses}
\end{table}

It is obvious from the table \ref{masses} that, higher is the rank of
the tensor, higher is the mass spectrum, and the possibility of
a finding these field in the LHC becomes more and more elusive. The
possibility for some signal of the $X$ field in LHC will comes from its
interaction with the dilaton field and this is analyzed in the next
section.

\section{Three form coupled to dilaton}

As we are going to explore here only the coupling between the three form and
the dilaton, we must consider only the relevant part of the action, which is
given by
\begin{align}
S_{X}&=\int d^{4}x\sum\limits_{n=0}^{\infty
}\sum\limits_{m=0}^{\infty }\int dye^{\phi /M^{3/2}}\nonumber\\
&\times[-2e^{4\sigma
}\chi ^{m}\chi ^{n}\eta ^{\alpha \beta }\eta ^{\mu \nu }\eta
^{\lambda \delta }\eta ^{\gamma \tau }Y_{\alpha \mu \lambda \gamma
}^{n}Y_{\beta \nu \delta \tau }^{m}\nonumber\\
&-8\left( m_{X}^{n}\right)
^{2}\eta ^{\alpha \beta }\eta ^{\mu \nu }\eta ^{\lambda \gamma
}X_{\alpha \mu \lambda }^{n}X_{\beta \nu \gamma }^{m}\chi ^{m}\chi
^{n}e^{4\sigma }].
\end{align}

When we expand the exponential of the dilaton, the first term will
give us the usual kinetic and mass term studied previously. The next
terms of the expansion are
\begin{align}
\exp({\phi/M^{3/2}})-1&=\left( M^{-3/2}\sum_{n=0}^{\infty }\phi ^{n}\left(
x\right) \frac{\psi ^{n}\left( y\right) }{\sqrt{r_{c}}}\right)\nonumber\\
&+\frac{1}{2!}
\left( M^{-3/2}\sum_{n=0}^{\infty }\phi ^{n}\left( x\right) \frac{\psi
^{n}\left( y\right) }{\sqrt{r_{c}}}\right) ^{2}+...
\end{align}

The terms beyond the first are highly suppressed by powers of the
plank mass, therefore we must consider only the first term, or
\begin{align}
&\exp({\phi /M^{3/2}})-1\sim M^{-3/2}\sum_{n=0}^{\infty }\phi ^{n}\left(
x\right) \frac{\psi ^{n}\left( y\right) }{\sqrt{r_{c}}}\nonumber\\
&=M^{-3/2}\left( \sqrt{%
kr_{c}}\phi ^{0}+\sum_{n=1}^{\infty }\sqrt{k}\frac{e^{2\sigma }}{%
e^{kr_{c}\pi }}\frac{J_{2}\left( z_{n}\right) }{J_{2}\left( x_{n}\right) }
\phi ^{n}\right).
\end{align}

The above term will give a coupling of the form $\phi X^{2}$. As we can see
from the complete solution of the dilaton and the three form, the
interaction of the massless fields will be suppressed by a factor of $\frac{1
}{M_{p}}$ and therefore is not considered. The interaction between a
massless dilaton, a massless two form and a massive three form is
like $\phi ^{0}X^{0}X^{n}$ and give us a null result because of the
orthogonality relations.

Therefore the interaction that posses massless dilatons are
irrelevant, and we must consider only the massive ones.
Taking in account these considerations and rearranging terms in the action,
we get for the cubic interactions
\begin{align}
S_{X}& \supset S_{cubic}=\sum\limits_{n,m=0}^{\infty }\sum_{p=1}^{\infty }%
\frac{M^{-3/2}}{\sqrt{r_{c}}}\int dye^{4\sigma }\chi ^{m}\left( y\right)
\chi ^{n}\left( y\right) \psi ^{p}\left( y\right)   \notag \\
& \times \int d^{4}x\phi ^{p}\left( x\right) [-2\eta ^{\alpha \beta }\eta
^{\mu \nu }\eta ^{\lambda \delta }\eta ^{\gamma \tau }Y_{\alpha \mu \lambda
\gamma }^{n}Y_{\beta \nu \delta \tau }^{m} \notag \\
& -8\left( m_{X}^{n}\right) ^{2}\eta ^{\alpha \beta }\eta ^{\mu \nu }\eta
^{\lambda \gamma }X_{\alpha \mu \lambda }^{n}X_{\beta \nu \gamma }^{m}].
\end{align}

From the effective action we can see that the coupling constant for these
interactions are given by
\begin{equation*}
\frac{M^{-3/2}}{\sqrt{r_{c}}}\int_{-\pi }^{+\pi }dye^{4\sigma }\chi ^{m}\chi
^{n}\psi ^{i}
\end{equation*}

As said before, due the large values of the mass for the KK modes, their signal must hardly be seen at LHC.
Therefore we concentrate in the more interesting case, that is the massless mode. Using now the solution for the
coefficients above we obtain, for the massless modes
\[
\frac{2kr_{c}}{M_{P}}e^{-5kr_{c}\pi }\int_{-\pi }^{+\pi }dye^{6\sigma }\frac{
J_{2}\left( z_{n}\right) }{J_{2}\left( x_{n}\right) }.
\]

The above integral was solved numerically using the software Mathematica, and the results for the first four KK modes are given in table \ref{coupling}

\begin{table}[h!]
\begin{center}
\begin{tabular}{|c|c|}
\hline
$n$ & Coupling constant\\ \hline
 1 & $0.00143676$  \\ \hline
 2 & $0.000788648$\\ \hline
 3 & $0.000413211$  \\ \hline
 4 & $0.000249329$ \\ \hline
\end{tabular}
\end{center}
\caption{Coupling constants in units of $Gev^{-1}$}
\label{coupling}
\end{table}
The values of the coupling constant in the above table are given in Gev$^{-1}$. This  show a rather interesting possibility:
Despite the fact that the massless mode is extremely suppressed, its coupling with the dilaton rises the possibility of a signal at LHC. A pointed in \cite{Mukhopadhyaya:2009gp}, the Lorenz structure of the interactions and masses involved will change the angular distribution of X fields produced. Furthermore, in the a Drell-Yan process, the production of X zero modes through , for example, the interaction of quark-antiquark pairs, may have different (but significant)  rates for sufficient integrated luminosity if we compare with the case of the Kalb-Ramond and gravity fields.

\section{Conclusions}

Extending earlier results about antisymmetric fields
\cite{Mukhopadhyaya:2007jn} we have shown here that the non-trivial
coupling between the dilaton and the antisymmetric field don't
affect its high suppression in our visible brane due to the warp
factor of the Randall-Sundrum scenario.

In the dilaton background we have studied new interactions between
the Kaluza-Klein modes of the dilaton field with the antisymmetric
field. These results are important from the phenomenological
viewpoint. The interactions that posses massless dilatons are
irrelevant, and we must consider only the massive ones. As we have
shown, higher order interactions of the massless fields are
suppressed by a factor of $\frac{1}{M_{p}}$ and therefore are not
considered. Due to the large mass of the massive modes, we considered
only the massless case. In fact, numerical computation of the important interaction is of order of Tev$^{-1}$ and can
give us signals of this field in the LHC's searches through Drell-Yan processes mediated by massive dilaton.

We thank Biswarup Mukhopadhyaya for helpful comments. The authors
would like to acknowledge the financial support provided by Funda\c c\~ao
Cearense de Apoio ao Desenvolvimento Cient\'\i fico e Tecnol\'ogico
(FUNCAP) and the Conselho Nacional de Desenvolvimento Cient\'\i fico e Tecnol\'ogico (CNPq).

This paper is dedicated to the memory of my wife  Isa\-bel Mara (R. R. Landim)

\end{document}